\documentclass{svjour2}
\usepackage{graphicx}
\begin{document}
\title{NMR in Superfluid A-like Phase of $^3$He Confined in
Globally Deformed Aerogel in Tilted Magnetic Field.}

\titlerunning{NMR in Superfluid A-like Phase}

\author{G.A. Baramidze         \and
        G.A. Kharadze.
}

\authorrunning{G.A. Baramidze \and G.A. Kharadze}

\institute{Andronikashvili Institute of Physics, 6 Tamarashvili
street, Tbilisi, 0177, Georgia\\
\email{g\_kharadze@yahoo.com}}

\date{Received: date / Accepted: date}

\maketitle

\begin{abstract}
NMR spectra in superfluid A-like phases confined in axially
deformed aerogel in presence of a magnetic field inclined with
respect to deformation axis is considered. The characteristic
features of dipole frequency shift in axially compressed and
axially stretched cases are compared. In particular, it is shown
that in axially stretched aerogel environment the stability region
of coherently spin precessing mode is rather narrow due to the
$U(1)$LIM effect. 
\end{abstract}

\section{Introduction}
\label{intro}

The NMR signature of spin-precessing modes of superfluid A-like
phase immersed in a uniaxially deformed aerogel attracts vivid
interest. An important information was collected from NMR spectra
of the A-like phase in axially compressed \cite{Kunimatsu,Sato}
and axially stretched \cite{Elbs} cases. In this situation the
orbital anisotropy axis $\hat{l}$ of the order parameter of
superfluid $^3$He-A is pinned either along cylindrical axis (in
axially compressed aerogel), or in the transverse plane (in
axially stretched aerogel).

The main body of information on the peculiarities of coherent spin
precession is contained in the transverse NMR frequency

\begin{equation}
\omega_{\bot}=\omega_L+\delta\omega_{\bot},
\label{eq:one}
\end{equation}
where the frequency shift $\delta\omega_{\bot}$ from the Larmor
value $\omega_L=gH$ is due to the action of the dipole-dipole
(spin-orbit) forces described by the potential

\begin{equation}
U_D=-\frac{1}{2}\chi\biggl(\frac{\Omega}{g}\biggr)^2(\hat{d}\cdot\hat{l})^2,
\label{eq:two}
\end{equation}
where $\hat{d}$ is the axis of magnetic anisotropy of $^3$He-A
order parameter. In a strong magnetic field $(\omega_L>>\Omega)$
the transverse NMR frequency shift

\begin{equation}
\delta\omega_{\bot}=-\frac{1}{S}\frac{\partial\bar{U}_D}{\partial\cos\beta},
\label{eq:three}
\end{equation}
where $\bar{U}_D$ denotes the time -averaged spin-orbit potentia,
$S=|\vec{S}|=\chi H/g$ is an equilibrium magnitude of spin density
and $\beta$ marks the tipping angle of $\vec{S}(t)$ precessing
about the direction of an applied magnetic field
$\vec{H}=H\hat{z}~(S_z=S\cos\beta)$.

In above-mentioned experiments (Refs.\cite{Kunimatsu,Sato,Elbs})
the magnetic field was directed along deformation axis of aerogel
samples filled with superfluid A-like phase and the information
about spin dynamics could be extracted from
$\delta\omega_{\bot}(\beta)$. In what follows we consider the case
where the orientation of $\vec{H}$ can be inclined by an angle
$\vartheta$ with respect to the axis of deformation of aerogel. In
this situation the observed  frequency shift
$\delta\omega_{\bot}(\beta,\vartheta)$ can be explored in
$(\beta,\vartheta)$ plane allowing to cover entire stability
region of coherently spin-precessing modes. The experiments using
the rotation of the magnetic field with respect to aerogel
deformation axis was, in particular, undertaken in
Ref.{\cite{Dmitriev}}. Our theoretical consideration contains some
suggestions for further experimental efforts.

In order to describe the configuration where the external magnetic
field $\vec{H}=H\hat{z}$ is inclined with respect to the axis of a
global deformation of aerogel we consider two coordinate frames:
an orbital frame $(\hat{\xi},\hat{\eta},\hat{\zeta})$ "attached"
to the cylindrical experimental cell (with $\hat{\zeta}$ oriented
along the cell axis) and the spin space frame
$(\hat{x},\hat{y},\hat{z})$. The orbital anisotropy axis is given
as:

\begin{equation}
\hat{l}=(\hat{\xi}\cos\phi_l+\hat{\eta}\sin\phi_l)\sin\lambda+\hat{\zeta}\cos\lambda.
\label{eq:four}
\end{equation}
In the spin-precessing regime the magnetic anisotropy axis
$\hat{d}(t)$ is rotating about an instantaneous orientation of
$\vec{S}(t)$, so that

\begin{equation}
\hat{d}(t)=\vec{R}(\alpha,\beta,\gamma)\hat{x}\bot
\vec{S}(t)=S\vec{R}(\alpha,\beta,\gamma)\hat{z}, \label{eq:five}
\end{equation}
where $(\alpha,\beta,\gamma)$ are Euler angles describing 3D
rotations in the spin space. In order to construct
$\delta\omega_{\bot}(\beta,\vartheta)$ we have to take into
account that for the orientation of $\vec{H}$ confined in the
$(\xi,\zeta)$ plane

\begin{eqnarray}
\hat{d}\cdot\hat{l}&=&d_x(\sin\lambda\cos\phi_l\cos\vartheta-\cos\lambda\sin\vartheta)+\nonumber\\
&+&d_y(\sin\lambda\sin\phi_l)+\\
&+&d_z(\sin\lambda\cos\phi_l\sin\vartheta+\cos\lambda\cos\vartheta).\nonumber
\label{eq:six}
\end{eqnarray}
In what follows we concentrate on two special orbital
configurations realized experimentally:

a) axially compressed aerogel with $\lambda=0$;

b) axially stretched aerogel with $\lambda=\pi/2$.

The results of our investigations are presented in the following
sequence. In Sec.2 the spin-precessing mode of the A-like phase in
axially compressed aerogel environment is considered. In Sec.3 the
spin-precessing mode of the A-like phase in axially stretched
aerogel environment is explored using the Volovik U(1)LIM model.
The conclusions are presented in Sec.4.

\section{Spin-precessing Mode of A-like Phase in Axially Compressed Aerogel.}
\label{sec:two}

In the considered long-ranged orbital configuration with
$\lambda=0$ the orbital phase $\phi_l$ is an irrelevant variable
and according to Eq.(6) the spin-orbit function

\begin{eqnarray}
f&=&(\hat{d}\cdot\hat{l})^2=(d_x\sin\vartheta-d_z\cos\vartheta)^2=\nonumber\\
&=& d_z^2+(d_x^2-d_z^2)\sin^2\vartheta-d_xd_z\sin 2\vartheta.
\label{eq:seven}
\end{eqnarray}
In an explicit form

\begin{eqnarray}
f(t)&=&\frac{1}{4}\Biggl\{2\sin^2\beta(1+\cos2\gamma)+\biggl [
-1+3\cos^2\beta+\nonumber\\
&+&\frac{1}{2}(1+\cos\beta)^2\cos2(\alpha+\gamma)+
\frac{1}{2}(1-\cos\beta)^2\cos2(\alpha-\gamma)-\nonumber\\
&-&\sin^2\beta (\cos 2\alpha+3\cos 2\gamma)\biggr
]\sin^2\vartheta+\\
&+&\sin\beta\bigl[2\cos\beta\cos\alpha+
(1+\cos\beta)\cos(\alpha+2\gamma)-\nonumber\\
&-&(1-\cos\beta)\cos(\alpha-2\gamma)\bigr]\sin
2\vartheta\Biggr\},\nonumber \label{eq:eight}
\end{eqnarray}
and at $S=S_{eq}=(\chi/g^2)\omega_L$ the phase
$\phi=\alpha+\gamma$ is a slow variable. As a result, $f(t)$ can
be decomposed in standard way \cite{Fomin} as

\begin{equation}
f(t)=\bar{f}+\tilde{f}(t),
\label{eq:nine}
\end{equation}
where the time-averaged (Van der Pol) part

\begin{equation}
\bar{f}=\frac{1}{2}\sin^2\beta+\frac{1}{4}\biggl[-1+3\cos^2\beta+
\frac{1}{2}(1+\cos\beta)^2\cos2\phi\biggr]\sin^2\vartheta,
\label{eq:ten}
\end{equation}
while the rapidly time-fluctuating contribution

\begin{eqnarray}
\tilde{f}(t)&=&\frac{1}{4}\Biggl\{2\sin^2\beta\cos
2\gamma+\Biggl[\frac{1}{2}(1-\cos\beta)^2\cos
2(\alpha-\gamma)-\nonumber\\
&-&\sin^2\beta(\cos 2\alpha+3\cos
2\gamma)\Biggr]\sin^2\vartheta+\Biggl [\sin
2\beta\cos\alpha+\nonumber\\
&+&\sin\beta(1+\cos\beta)\cos(\alpha+2\gamma)-\nonumber\\
&-&\sin\beta(1-\cos\beta)\cos(\alpha-2\gamma)\Biggr]\sin
2\vartheta\Biggr\}, \label{eq:afterten}
\end{eqnarray}
gives life to the small amplitude high-frequency spin fluctuations
superimposed on the coherently spin-precessing modes emerging from
the Van der Pol picture. As follows from Eq.(\ref{eq:ten}), in
orbital state with $\lambda=0$

\begin{equation}
\bar{U}_D=-\frac{1}{2}\chi\biggl(\frac{\Omega}{g}\biggr)^2 \biggl[
\frac{1}{2}\sin^2\beta+\frac{1}{4}\biggl(-1+3\cos^2\beta+
\frac{1}{2}(1+\cos\beta)^2\cos2\phi\biggr)\sin^2\vartheta\biggr].
\label{eq:eleven}
\end{equation}
The phase $\phi$ entering in Eq.(\ref{eq:eleven}) can be fixed by
minimizing $\bar{U}_D$ at $\phi=\phi_{st}=(0,\pi)$:

\begin{equation}
\bar{U}_D=-\frac{1}{2}\chi\biggl(\frac{\Omega}{g}\biggr)^2 \biggl[
\frac{1}{2}\sin^2\beta+\frac{1}{8}(-1+2\cos\beta+7\cos^2\beta)\sin^2\vartheta\biggr],
\label{eq:twelve}
\end{equation}
and according to Eq.(\ref{eq:three}) the dipole frequency shift

\begin{equation}
\delta\omega_{\bot}(\beta,\vartheta)=\biggl(\frac{\Omega^2}{2\omega_L}\biggr)
\biggl[-\cos\beta+\frac{1}{4}(1+7\cos\beta)\sin^2\vartheta
\biggr].
\label{eq:thirteen}
\end{equation}
The stability criterion (the concavity of $\bar{U}_D$ with respect
to $\cos\beta$) is

\begin{equation}
\frac{\partial^2\bar{U}_D}{\partial(\cos\beta)^2}>0~\Rightarrow~\cos^2\vartheta>\frac{3}{7},
\label{eq:foureteen}
\end{equation}
which is in accordance with the conclusion found in
Ref.\cite{Gurgenishvili}.

In order to reach a contact with the description of the coherently
spin-precessing state in terms of the magnon BEC \cite{Bunkov_1}
one has to take into account that an exited spin-dynamical mode
with fixed frequency and phase  is characterized by the Bose
quasiparticles - magnons with density

\begin{equation}
n_M=|\psi|^2=S-S_z=S(1-\cos\beta),
\label{eq:fifteen}
\end{equation}
so that Eq.(\ref{eq:twelve}) for $\bar{U}_D$ can be transcribed as

\begin{equation}
\bar{U}_D(\beta,\vartheta)=-\frac{1}{2}\chi\biggl(\frac{\Omega}{g}
\biggr)^2\biggl[\sin^2\vartheta+\frac{|\psi|^2}{S}\cos 2\vartheta-
\frac{7}{8}(\cos^2\vartheta-\cos^2\vartheta_o)\frac{|\psi|^4}{S^2}\biggr],
\label{eq:sixteen}
\end{equation}
where $\cos^2\vartheta_o=3/7$. From Eq.(\ref{eq:sixteen}) it
follows that magnon-magnon interaction potential

\begin{equation}
U_M(\vartheta)=\frac{7}{4^2}\chi\biggl(\frac{\Omega}{g} \biggr)^2
(\cos^2\vartheta-\cos^2\vartheta_o), \label{eq:seventeen}
\end{equation}
and the magnon-magnon repulsion regime
($\cos^2\vartheta>\cos^2\vartheta_o$) reproduces the stability
criterion (\ref{eq:foureteen}) of coherent spin-precessing mode,
as expected.

\begin{figure}
\includegraphics[width=0.75\textwidth]{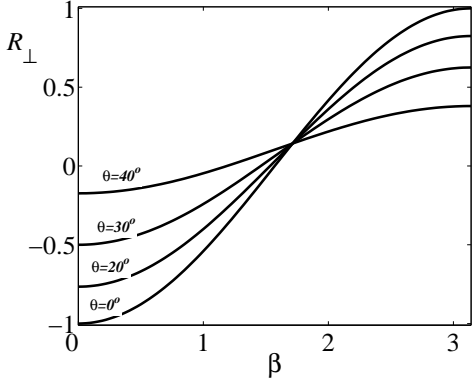}
\caption{Dependence of normalized dipole frequency shift
$R_{\bot}(\beta,\vartheta)=\delta\omega_{\bot}(\beta,\vartheta)
/(\Omega^2/2\omega_L)$ on $\beta$ at various inclination angles
$\vartheta$.} \label{fig:1}
\end{figure}

Now, the dipole frequency shift for an axially compressed aerogel
(at $\lambda=0$) can be represented as

\begin{eqnarray}
\delta\omega_{\bot}(\beta,\vartheta)&=&\biggl(\frac{\Omega^2}{2\omega_L}
\biggr)\biggl[-\cos
2\vartheta+\frac{7}{4}(\cos^2\vartheta-\cos^2\vartheta_o)(1-\cos\beta)\biggr]=\nonumber\\
&=&\frac{1}{4}\biggl(\frac{\Omega^2}{2\omega_L}
\biggr)[\sin^2\vartheta-7(\cos^2\vartheta-\cos^2\vartheta_o)\cos\beta)].
\label{eq:eighteen}
\end{eqnarray}
In Ref.\cite{Dmitriev} it was shown, in particular, that in linear
cw NMR regime ($\beta \rightarrow 0$) the experimental data are
well described by the formula

\begin{equation}
\delta\omega_{\bot}(0,
\vartheta)=-\biggl(\frac{\Omega^2}{2\omega_L}\biggr)\cos
2\vartheta. \label{eq:ninteen}
\end{equation}
On the other hand, in the pulsed NMR case for $\vartheta=\pi/2$ it
was found that the dipole frequency shift is described by the
Brinkman-Smith formula

\begin{equation}
\delta\omega_{\bot}(\beta,
\pi/2)=\biggl(\frac{\Omega^2}{2\omega_L}\biggr)\frac{1}{4}(1+3\cos
\beta). \label{eq:twenty}
\end{equation}
It should be remarked that according to Ref.\cite{Sato} the
measured $\beta$-dependence of dipole frequency shift at
$\vartheta=0$

\begin{equation}
\delta\omega_{\bot}(\beta,
0)=-\biggl(\frac{\Omega^2}{2\omega_L}\biggr)\cos \beta.
\label{eq:twentyone}
\end{equation}
All of the mentioned experimental results are in accordance with
general Eq.(\ref{eq:eighteen}). At the same time, the answer for
an axially compressed aerogel case (with $\lambda=0$) contains
some hints to stimulate further experimental efforts. In
addressing Eq.(\ref{eq:eighteen}) we see that at the edge of the
stability of coherently spin-precessing mode realized at the
magnetic field tilting angle $\vartheta=\vartheta_o$ the dipole
frequency shift should be independent of spin-tipping angle
$\beta$:

\begin{equation}
\delta\omega_{\bot}(\beta,
\vartheta_o)\equiv\frac{1}{7}\biggl(\frac{\Omega^2}{2\omega_L}\biggr).
\label{eq:twentytwo}
\end{equation}
On the other hand, at $\cos\beta_o=-1/7$ the dipole frequency
shift should be independent of the magnetic field inclination
angle $\vartheta$:

\begin{equation}
\delta\omega_{\bot}(\beta_o,
\vartheta)\equiv\frac{1}{7}\biggl(\frac{\Omega^2}{2\omega_L}\biggr).
\label{eq:twentythree}
\end{equation}
This behaviour of $\delta\omega_{\bot}(\beta,\vartheta)$, given by
Eq.(\ref{eq:eighteen}), is clearly seen in Fig.\ref{fig:1}.

It is to be mentioned that Eq.(\ref{eq:eighteen}) describing the
dipole frequency shift $\delta\omega_{\bot}(\beta,\vartheta)$ is
applicable to $^3$He-A confined in the narrow gap with orbital
axis $\hat{l}$ pinned to the normal of parallel plates. The
corresponding experiments \cite{Freeman} were performed in the
case of $\vec{H}\parallel\hat{l}~(\vartheta=0)$. Unfortunately no
attempt to explore the $\vartheta$-dependence of
$\delta\omega_{\bot}(\beta,\vartheta)$ was undertaken at that
time.

\section{Spin-precessing Mode of A-like Phase in Axially Stretched Aerogel.}
\label{sec:three}

Now, we proceed to an axially stretched (radially squeezed)
aerogel case corresponding to the orbital configuration with
$\lambda=\pi/2$. In referring to Eq.(6) it is readily seen that in
this situation

\begin{eqnarray}
(\hat{d}\hat{l})^2&=&d_x^2\cos^2\phi_l\cos^2\vartheta+d_y^2\sin^2\phi_l+
d_z^2\cos^2\phi_l\sin^2\vartheta+\nonumber\\
&+&d_xd_y\sin 2\phi_l\cos\vartheta+d_xd_z\cos^2\phi_l\sin
2\vartheta+d_yd_z\sin 2\phi_l\sin\vartheta. \label{eq:twentyfour}
\end{eqnarray}
According to the $U(1)$LIM model \cite{Elbs} for the fully
randomized $\phi_l$

\begin{equation}
\langle\sin^2\phi_l\rangle=\langle\cos^2\phi_l\rangle=\frac{1}{2},~~~~~~
\langle\sin 2\phi_l\rangle=0, \label{eq:twentyfive}
\end{equation}
and it is concluded that in an axially stretched aerogel

\begin{eqnarray}
(\hat{d}\hat{l})^2&=&\frac{1}{2}(d_x^2\cos^2\vartheta+d_y^2+
d_z^2\sin^2\vartheta+d_xd_z\sin 2\vartheta)=\nonumber\\
&=&\frac{1}{2}(\hat{d}^2(t)-(d_x\sin\vartheta-d_z\cos\vartheta)^2)
=-\frac{1}{2}f(t)+\textrm{const}, \label{eq:twentysix}
\end{eqnarray}
where $f(t)$ is given by Eq.(8). In constructing the time-averaged
(Van der Pol) dipole-dipole potential

\begin{equation}
\langle\bar{U}_D\rangle=\frac{1}{4}\chi\biggl(\frac{\Omega}{g}\biggr)^2\bar{f},
\label{eq:twentyseven}
\end{equation}
we see that according to Eq.(\ref{eq:ten}) it is minimized at
$\phi=\pi/2$ and for the dipole frequency shift
$\delta\omega_{\bot}$ the following answer is obtained

\begin{equation}
\delta\omega_{\bot}(\beta,\vartheta)=\frac{1}{2}\biggl(\frac{\Omega^2}{2\omega_L}
\biggr)\biggl[\cos\beta+\frac{1}{4}(1-5\cos\beta)\sin^2\vartheta
\biggr]. \label{eq:twentyeight}
\end{equation}
In particular,

\begin{equation}
\delta\omega_{\bot}=\biggl(\frac{\Omega^2}{2\omega_L}
\biggr)\left\{
\begin{array}{cc}\frac{1}{2}\cos\beta,&\vartheta=0\\
\\
\frac{1}{8}(1-\cos\beta),&\vartheta=\pi/2.\end{array}\right.
\label{eq:twentnine}
\end{equation}

\begin{figure}[t]
\includegraphics[width=0.75\textwidth]{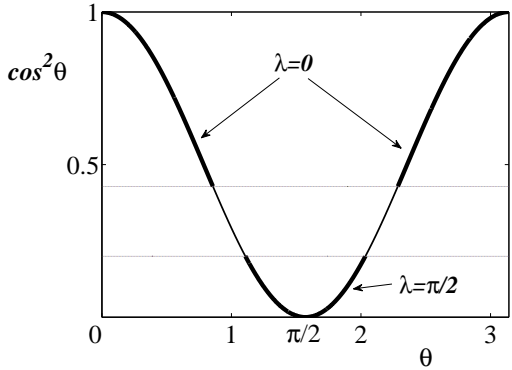}
\caption{Stability regions of coherently spin-precessing modes in
axially compressed ($\lambda=0$) and axially stretched
($\lambda=\pi/2$) aerogel.} \label{fig:2}
\end{figure}

In the magnon BEC representation

\begin{equation}
\langle\bar{U}_D\rangle=\frac{1}{4}\chi\biggl(\frac{\Omega}{g}\biggr)^2\biggl[
\frac{|\psi|^2}{S}\cos^2\vartheta-\frac{5}{8}\frac{|\psi|^4}{S^2}
(\cos^2\vartheta-\cos^2\vartheta_o)\biggr], \label{eq:thirty}
\end{equation}
where $\cos^2\vartheta_o=1/5$. Consequently, the magnon-magnon
interaction potential

\begin{equation}
U_M({\vartheta})=-\frac{5}{2^5}\chi\biggl(\frac{\Omega}{g}\biggr)^2
(\cos^2\vartheta-\cos^2\vartheta_o), \label{eq:thirtyone}
\end{equation}
and it is repulsive at $\cos^2\vartheta<\cos^2\vartheta_o$. The
stability region of spin-precessing mode in the case of an axially
stretched aerogel environment $(\lambda=\pi/2)$, due to the
$U(1)$LIM effect, is confident in a narrow gap

\begin{equation}
0\leq\cos^2\vartheta<\frac{1}{5}, \label{eq:thirtytwo}
\end{equation}
being essentially different from axially compressed aerogel case
with $\lambda=0$ (see Eq.(\ref{eq:foureteen})and Fig.\ref{fig:2}).
This means, in particular, that the spin-precessing mode realized
at $\vartheta=0$ (see Eq.(\ref{eq:twentnine})) and discovered
experimentally \cite{Elbs} is an unstable state. It would be
desirable to support this conclusion by observation of the fast
decay of FIS appropriate to an unstable spin-precessing mode. In
an open geometry such test of the instability of spin-precession
in $^3$He-A for the Leggett orbital configuration have been
performed in Refs.\cite{Borovik,Bunkov_2} confirming the
prediction made in Ref.\cite{Fomin_1}. On the other hand, the
expectation for the transverse orientation of
$\vec{H}(\vartheta=\pi/2)$, see Eq.(\ref{eq:twentnine}), which is
in the limits of stability region (\ref{eq:thirtytwo}), is still
to be confirmed experimentally.

Returning back to the $\beta$-representation we find that in an
axially stretched aerogel case

\begin{eqnarray}
\delta\omega_{\bot}(\beta,\vartheta)&=&\frac{1}{2}\biggl(\frac{\Omega^2}{2\omega_L}\biggr)
\biggl[\cos^2\vartheta-\frac{5}{4}(\cos^2\vartheta-\cos^2\vartheta_o)(1-\cos\beta)
\biggr]=\nonumber\\
&=&\frac{1}{8}\biggl(\frac{\Omega^2}{2\omega_L}\biggr)[\sin^2\vartheta+
5(\cos^2\vartheta-\cos^2\vartheta_o)\cos\beta].
\label{eq:thirtythree}
\end{eqnarray}

\section{Conclusion}
\label{concl}

In this article the specific features of the NMR spectra of the
A-like phase of superfluid $^3$He in uni-axially deformed aerogel
environment is considered in presence of an inclined magnetic
field with respect to global deformation axis. Two special orbital
configurations, relevant to experimentally investigated cases, are
analyzed: a) an axially compressed aerogel with orbital anisotropy
axis $\hat{l}$ pinned to deformation axis; b) an axially stretched
aerogel with $\hat{l}$ confined in transverse plane with respect
to deformation axis. The later case is treated in terms of
$U(1)$LIM model. In the mentioned cases the frequency shift
$\delta\omega(\beta,\vartheta)$ from the Larmor value
$\omega_L=gH$ is constructed. Among other things, the stability
limits of coherently spin-precessing modes with respect to
magnetic field inclination angle $\vartheta$ are found. In
particular, it is shown that in an axially stretched aerogel
environment the $U(1)$LIM effect reduces the stability region to a
narrow gap around  $\vartheta=\pi/2$. Some suggestions for further
experimental efforts are given.

\end{document}